\documentclass[a4paper,
              onecolumn,
              11pt,
              accepted=2023-04-24
              ]{quantumarticle}
\pdfoutput=1
\usepackage[numbers,sort&compress]{natbib}
\usepackage[utf8]{inputenc}
\usepackage[english]{babel}
\usepackage{braket}
\usepackage[T1]{fontenc}
\usepackage{amsmath}
\usepackage{subfigure}
\usepackage{hyperref}
\usepackage{xcolor}
  \definecolor{codegreen}{rgb}{0,0.6,0}
  \definecolor{codegray}{rgb}{0.5,0.5,0.5}
  \definecolor{codepurple}{rgb}{0.58,0,0.82}
  \definecolor{backcolour}{rgb}{0.95,0.95,0.92}
\usepackage{listings}
  \lstdefinestyle{mystyle}{
    backgroundcolor=\color{backcolour},   
    commentstyle=\color{codegreen},
    keywordstyle=\color{magenta},
    numberstyle=\tiny\color{codegray},
    stringstyle=\color{codepurple},
    basicstyle=\ttfamily\footnotesize,
    breakatwhitespace=false,
    breaklines=true,
    captionpos=b,
    keepspaces=true,
    numbersep=5pt,
    showspaces=false,
    showstringspaces=false,
    showtabs=false,
    tabsize=2,
    escapechar=\%
}
\newcommand{\ourpackagename}[1]{\textsf{Inflation}}

\begin{document}

\title{Inflation: a Python library for classical and quantum causal compatibility}

\author{Emanuel-Cristian Boghiu}
\affiliation{ICFO – Institut de Ciencies Fotoniques, The Barcelona Institute of Science and Technology, 08860 Castelldefels (Barcelona), Spain}
\email{cristian.boghiu@icfo.eu}
\orcid{0000-0003-0513-9004}

\author{Elie Wolfe}
\affiliation{Perimeter Institute for Theoretical Physics, 31 Caroline St. N., Waterloo, Ontario, Canada, N2L 2Y5}
\email{ewolfe@pitp.ca}
\orcid{0000-0002-6960-3796}

\author{Alejandro Pozas-Kerstjens}
\affiliation{Instituto de Ciencias Matem\'aticas (CSIC-UAM-UC3M-UCM), 28049 Madrid, Spain}
\email{physics@alexpozas.com}
\orcid{0000-0002-3853-3545}

\maketitle

\begin{abstract}
    We introduce \ourpackagename{}, a Python library for assessing whether an observed probability distribution is compatible with a causal explanation.
    This is a central problem in both theoretical and applied sciences, which has recently witnessed significant advances from the area of quantum nonlocality, namely, in the development of inflation techniques.
    \ourpackagename{} is an extensible toolkit that is capable of solving pure causal compatibility problems and optimization over (relaxations of) sets of compatible correlations in both the classical and quantum paradigms.
    The library is designed to be modular and with the ability of being ready-to-use, while keeping an easy access to low-level objects for custom modifications.
\end{abstract}

\section{Motivation}
One of the main challenges in any scientific discipline is identifying which are the causes behind some observed correlations.
Is a vaccine effective against a disease? Does raising salaries encourage spending? Is the increase of carbon dioxide in the atmosphere responsible for the increase in the average temperature of the Earth? These questions and their like can all be phrased and analyzed using the tools of causal inference (CI)~\cite{causalBook}.
However, despite the wide relevance of causal inference, current CI algorithms involving latent variables are typically incapable of analyzing structures with more than a small number of nodes~\cite{geiger1995,tian2002,garcia2005,garcia2014,lee2017}.

The field of quantum nonlocality~\cite{BellReview} has recently put its attention in causality, in light of the fact that Bell's theorem~\cite{BellTheorem} can be understood in terms of the compatibility of probability distributions with a given causal structure~\cite{wood2015,chaves2015}.
This view has propelled the study of quantum correlations beyond the traditional bipartite scenario (see, e.g., \cite{bilocal_branciard_prl,bilocal_branciard_pra,Fritz_2012,fraser2018,vanhimbeeck2019}, and the review \cite{networkReview}), and the development of techniques for characterizing the quantum and classical probability distributions that can be generated in such causal scenarios \cite{scalarextension,covariancematrices1,covariancematrices2,luo2018,RenouFinner2019}.
A particularly successful tool is the inflation method~\cite{wolfe2019,wolfe2021,gisin2020}, which consists of a series of increasingly strict necessary conditions that can be tested via linear or semidefinite programming.
Despite its broad applicability within and outside the field of quantum nonlocality, available implementations of the inflation technique are typically limited in terms of the type of causal structures it applies to, or in the type of inflations considered (see, e.g., \cite{fullnn,pozas2022proofs}).
This means that researchers must code their own programs every time they seek to analyze a different structure or try a different solution, adding an extra level of difficulty to the application of the technique.

Here we present \ourpackagename{}~\cite{code}, an open-source library, written in Python, that implements the inflation framework for causal compatibility.
It allows both for solving feasibility problems (i.e., answering the question \textit{``can I generate this distribution in this causal structure?''}) and for bounding optimal values of functionals over distributions compatible with a causal structure.
These include all network scenarios considered in the field of quantum nonlocality~\cite{networkReview}, and structures that have recently gained attention in that field, such as the so-called instrumental scenario~\cite{vanhimbeeck2019}.

Currently, \ourpackagename{} implements the quantum inflation hierarchy of Ref.~\cite{wolfe2021}.
This means that the main focus is the characterization of distributions generated by measuring quantum systems.
However, the library can also be used, by setting the corresponding flags, for assessing compatibility with distributions generated when all the latent nodes represent sources of classical shared randomness (thereby extending the ideas in \cite{baccari2017,Steeg2011}), and when all the latent nodes represent sources of quantum entanglement and at the same time all parties are correlated through a global source of classical shared randomness.

This paper presents the package and illustrates simple use cases, which are extended in the library documentation.
It is structured as follows: in Sec.~\ref{sec:inflation} we provide a brief description of the theoretical ideas behind inflation and point to the relevant literature.
In Sec.~\ref{sec:library} we show how to get started with the library and describe its main components and features.
In Secs.~\ref{sec:main} and \ref{sec:other} we demonstrate with code snippets the different types of problems that can be addressed with \ourpackagename{}.
We discuss further software details and library information in Sec.~\ref{sec:additional} including future development, contribution guidelines, and planned maintenance and support, and we provide some concluding remarks in Sec.~\ref{sec:conclusions}.

\section{The inflation framework}
\label{sec:inflation}
The aim of this section is to provide a brief description of the general ideas behind inflation.
For the reader interested in the details of the different variants, we refer to the original publications~\cite{wolfe2019,navascues2020,wolfe2021}.

\begin{figure}
    \centering
    \subfigure[\label{fig:bell}]
    {\begin{minipage}[t]{0.20\textwidth}
        \vspace*{-3cm}
          \centering\includegraphics[width=0.9\textwidth]{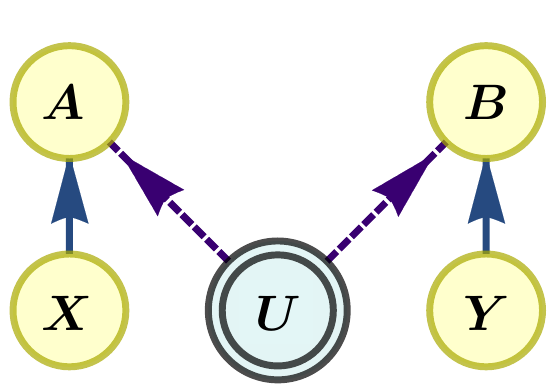}
      \end{minipage}
      }
      \hspace*{0.15cm}
    \subfigure[\label{fig:triangle}]
    {\begin{minipage}[t]{0.37\textwidth}
          \hspace*{0.9cm}\includegraphics[width=0.9\linewidth]{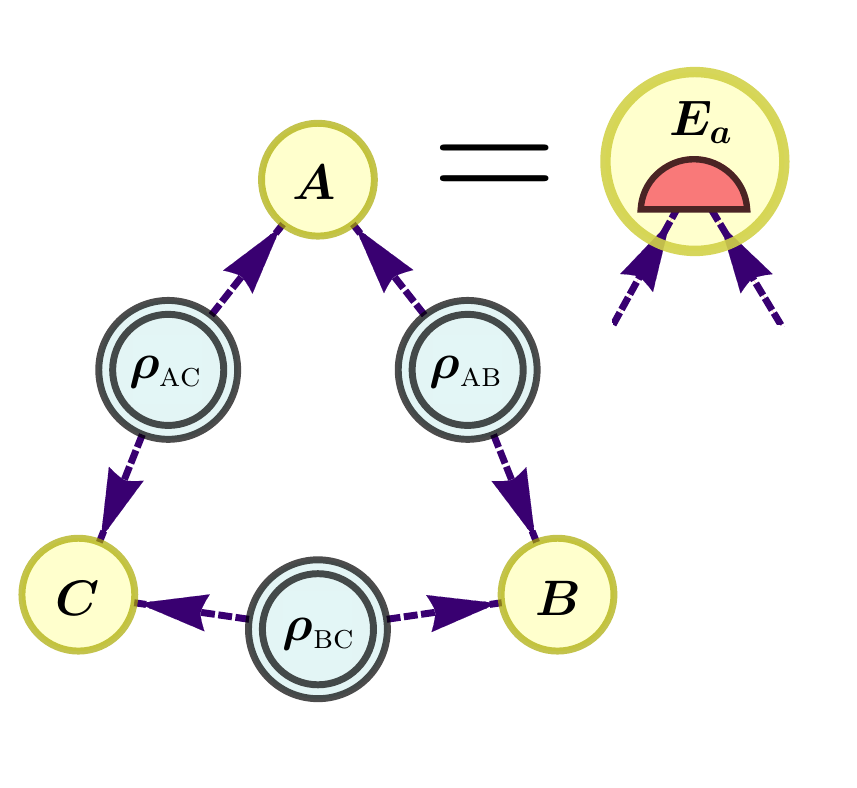}
      \end{minipage}
      }
    \subfigure[\label{fig:triangleinflation}]
    {\begin{minipage}[t]{0.37\textwidth}
          \hspace*{0.8cm}\includegraphics[width=0.9\textwidth]{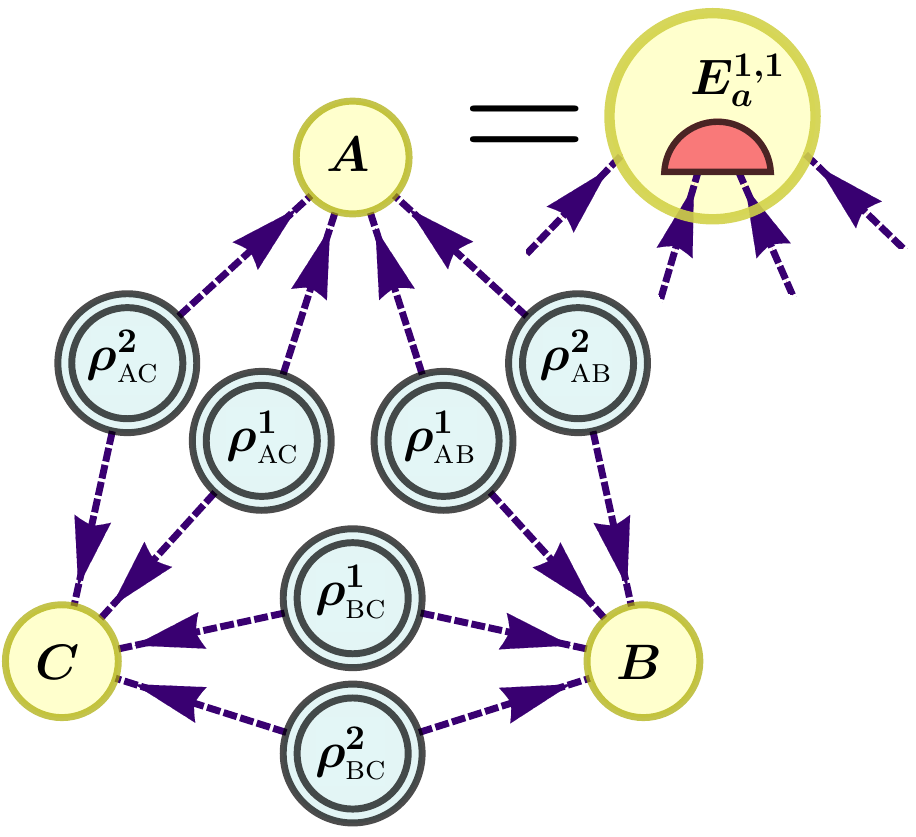}
      \end{minipage}
      }
    \caption{\protect{\subref{fig:bell}} The bipartite Bell scenario written as a DAG.
    The yellow circles denote visible nodes representing random variables, while the blue circles denote a latent node representing a physical system.
    This DAG represents the scenario where two parties perform measurements denoted by $x\sim X$ and $y \sim Y$, respectively, on shares of a bipartite physical system distributed to them.
    The results of the measurements are $a\sim A$ and $b \sim B$.
    \protect{\subref{fig:triangle}} The triangle scenario, where three parties perform (in this case, fixed) measurements on shares of bipartite physical systems. \protect{\subref{fig:triangleinflation}} The second-order quantum inflation of the triangle scenario.
    Each of the sources is duplicated, and each party has a choice of the shares in which to perform their measurements.
    Whereas NPA-based convex relaxations of scenario \protect{\subref{fig:triangle}} are indistinguishable from those of single common cause scenarios, the underlying symmetries in \protect{\subref{fig:triangleinflation}} allow to constrain correlations beyond the common-cause scenario.
    }
    \label{fig:inflation}
\end{figure}

Inflation is a general framework for analyzing correlations that can be generated in causal scenarios.
These scenarios feature latent nodes, corresponding to physical systems (sources of shared classical randomness, quantum states, or states of more general physical theories), and visible nodes, which represent random variables that describe the outcomes of measurements performed on such systems.
The nodes are connected by arrows that denote which systems are sent to which parties. In order to avoid causal paradoxes, like the outcome of a measurement determining which measurement is performed in the first place, the paths created by following the arrows must not contain closed loops. These graphs with directed arrows and no loops are known as directed acyclic graphs, or DAGs.
Examples of such DAGs can be found in Figs.~\ref{fig:inflation} and \ref{fig:instrumental}.

In order to characterize the distributions that can be generated in a particular causal scenario, inflation considers the \textit{gedankenexperiment} where one has access to multiple copies of the physical systems (recall, the latent nodes in the corresponding DAG) and operations (the visible nodes) used in it.
By connecting these elements in such a way that the parents of a copy of a visible node are copies of the parents of that node in the original scenario, one constructs \textit{inflation scenarios} where the characterization of compatible distributions is simpler than in the original scenario.
Furthermore, constraints satisfied by compatible distributions can be translated into necessary constraints of distributions compatible with the original scenario.
The distributions compatible with inflated scenarios satisfy two main properties.
First, they are highly symmetric, due to the fact that the elements of the inflation scenarios are copies of the elements of the original scenario.
Namely, distributions compatible with inflations are invariant under permutations of copies of a same latent node.
Second, when marginalized over sets of nodes that reproduce parts of the original scenario, they coincide with the corresponding marginals of the distribution in the original scenario.
These two properties can be encoded by means of linear equalities and inequalities.
Therefore, probability distributions that can be generated in inflations can be characterized by means of linear programming in the case where the physical systems are classical \cite{wolfe2019} or described by a generalized physical theory \cite{gisin2020}, or with hierarchies of semidefinite programs in the case where the systems measured by the parties are quantum mechanical \cite{wolfe2021}.
In all these cases, efficient algorithms exist that allow to solve the corresponding problems with standard computing resources.

Inflation is typically used for two families of problems: optimizing quantities over distributions compatible with some causal scenario, and determining whether a particular distribution admits a realization in a given scenario.
For the former, inflation provides bounds to the actual optima (which can be nevertheless arranged in monotonic sequences, see e.g. Refs.~\cite{navascues2020,wolfe2021}).
For the latter, finding an inflation where all the symmetry and marginal constraints implied cannot be satisfied is a proof that the original premise, namely that the distribution under scrutiny admitted a realization in the causal structure, is false.
Proving the opposite, namely that the original distribution admits a realization in the original causal scenario, requires finding an explicit realization (in terms of classical shared randomness, quantum systems, or elements of a generalized probabilistic theory, as corresponding) of the distribution.
Exploring such constructions is a task outside the scope inflation, and thus of this library.

\ourpackagename{} currently implements the quantum inflation hierarchy of Ref.~\cite{wolfe2021}. This is, it considers that all sources in the scenario distribute quantum systems with an unbounded dimension. As such, whenever a correlation can be achieved using positive operator values measurements (POVMs) the same correlation can also be realized by projective-valued measures (PVMs), as any POVM can be dilated to a PVM through Naimark's dilation theorem. As such, \ourpackagename{} models all measurements as projective without loss of generality. This leads to a hierarchy that is monotonic (so each level is, at least, as constraining as the previous one), but that until now it has not been proven to converge except in concrete scenarios \cite{ligthart2022}. There also exist alternative hierarchies that have been proven to converge \cite{ligthart2021}, although in a non-monotonic way.

\section{The library}
\label{sec:library}
\subsection{Requirements and installation}
\ourpackagename{} is a Python library that can be installed on Mac, Windows, and Linux operating systems via \texttt{pip} by executing the instruction below at a command line.
\begin{lstlisting}[language=Python, captionpos=t]
    pip install inflation
\end{lstlisting}
The core requirements of \ourpackagename{} are NumPy \cite{numpy}
(used for general numerical procedures), SymPy \cite{sympy} (used to make the input format more user-friendly), and SciPy \cite{scipy} (used for handling sparse matrices).
It can also use Numba \cite{numba} as a just-in-time compiler to speed up core calculations.
For solving the generated relaxations, the library uses the MOSEK Fusion API \cite{mosek} to solve the linear and semidefinite programming problems.
It also allows for writing the problem in a human-readable form as a comma-separated values file, to a MATLAB-compatible .mat file, or to SDPA data format for further manipulation in other interfaces such as Yalmip \cite{yalmip}.

To test installation, one can run the following.
\begin{lstlisting}[language=Python, captionpos=t, basicstyle= \fontsize{8}{10}\selectfont\ttfamily]
import inflation
inflation.about()
\end{lstlisting}
These lines print basic information about the version of \ourpackagename{} installed and the versions of installed dependencies.
\begin{lstlisting}[language=Python, captionpos=t, basicstyle= \fontsize{8}{10}\selectfont\ttfamily]
# Inflation: Implementations of the Inflation Technique for Causal Inference
# ================================================================================
# Authored by: Emanuel-Cristian Boghiu, Elie Wolfe and Alejandro Pozas-Kerstjens

# Inflation Version:        0.1
# 
# Core Dependencies
# -----------------
# NumPy Version:	1.23.1
# SciPy Version:	1.8.1
# SymPy Version:	1.11.1
# Numba Version:	0.56.2
# Mosek Version:	10.0.20
#
# Python Version:	3.10.4
# Platform Info:	Windows (AMD64)
\end{lstlisting}

The source code for \ourpackagename{} is hosted on GitHub at

\centerline{\href{https://github.com/ecboghiu/inflation}{https://github.com/ecboghiu/inflation}}
\noindent and is distributed with an open-source software license: GNU GPL version 3.0.
More details about the software, packaging information, and guidelines for contributing to \ourpackagename{} are included in Sec.~\ref{sec:additional}.
\subsection{Components}
There are two main layers in \ourpackagename{}.
First, the basic characterization of the causal scenario and its desired inflation are stored in an \texttt{InflationProblem}.
This includes the DAG describing the causal structure, the number of inputs and outputs of each of its visible nodes, and the number of copies of each latent node in the desired inflation.
If the causal structure furthermore contains visible-to-visible connections, then the procedures in, e.g., \cite[Sec. V]{navascues2020}, \cite[Sec. V]{wolfe2021} and \cite[Sec. IV.C]{ligthart2021} are executed in order to find the network and suitable constraints that generate the same distributions as the original causal structure.

The second layer takes the characterization provided by an \texttt{InflationProblem} object and sets up and solves the compatibility or optimization problem of interest.
Currently, the library only supports the quantum inflation hierarchy described in Ref.~\cite{wolfe2021} via the \texttt{InflationSDP} object.
However, the implementation also supports the analysis of distributions generated from classical latent nodes by imposing suitable commutation constraints (see Ref.~\cite{baccari2017} and \cite[Sec. VI]{wolfe2021}).
Sec.~\ref{sec:other:classical} showcases how this can be achieved in \ourpackagename{}.

In the quantum inflation hierarchy, the characterization of the set of probability distributions is given by a list of operators, in the spirit of the Navascu\'es-Pironio-Ac\'in (NPA) hierarchy~\cite{npa1,npa2,pna}.
This is input to the \texttt{InflationSDP} object via the function \texttt{generate\_relaxation()}, which admits either a generic list-of-lists notation for arbitrary lists of operators, or string-based notations for operator sets that are routinely used in the literature: \texttt{npa\#} for the sets describing the NPA hierarchy (denoted as $\mathcal{T}_n$ in Ref.~\cite{npa1} and as $\mathcal{S}_n$ in Ref.~\cite{npa2}), \texttt{local\#} for the so-called local levels (denoted as $\mathcal{L}_n$ in \cite[App. C]{wolfe2021}, see also Ref.~\cite{moroder2013}), and \texttt{physical\#} for the sets of operators of bounded length whose expectation value is non-negative for any quantum state.

The function \texttt{generate\_relaxation()} automatically imposes the equality constraints that are derived from the invariance of the inflation under permutation of copies of a same original element (this is, it imposes the constraints described in \cite[Eq. (10)]{wolfe2021}).
Furthermore, it identifies which marginals of distributions in the inflation must coincide with marginals of distributions in the original scenario.
After this, the user can specify either a probability distribution over the visible nodes for a feasibility problem (i.e., to determine whether the distribution can be identified as incompatible with the causal structure) using the function \texttt{set\_values()}, or a combination of operators whose expectation value will be optimized, by using the function \texttt{set\_objective()}.
When using \texttt{set\_values()}, the user can choose to set also the so-called linearized polynomial constraints, which constrain the set of compatible distributions further at the expense of obtaining certificates with more limited applicability~\cite{AlexThesis,pozas2022proofs}.

\subsection{Reductions of the feasible region}
In general, the fact that one must consider multiple copies of the elements in the original causal structure leads to a large computational load when storing and solving the relevant problems.
\ourpackagename{} implements a number of additional constraints not included in the original definitions of the hierarchies, either automatically or at the user's choice, that give a tighter relaxation for the same level of computational resources.
These constraints are:
\paragraph{Non-negativity of physical moments.} This is a feature only relevant for the implementations of inflation that characterize distributions generated by measuring quantum systems.
Implementations of quantum inflation require the use of the NPA hierarchy \cite{npa1} in order to assess whether a compatible inflation exists.
The main object in the NPA hierarchy are the so-called moment matrices, whose rows and columns are indexed by products of (a priori unknown) projection operators.
Each cell of the moment matrix contains the expectation of the product of the operators in the row and the operators in the column under an also unknown quantum state.
Despite all elements being unknown a priori, one can derive constraints for some of them in certain situations.
For instance, it is known that eigenpotent operators have a non-negative expectation value under any possible quantum state, and thus these non-negativity constraints are always imposed in \ourpackagename{}.
In fact, using operator products which have a non-negative expectation value under any quantum state in the generating set for \texttt{InflationSDP} leads to drastic reductions in problem size for certain problems, as we explicitly show in Sec.~\ref{sec:main:feas}.

\paragraph{Sandwich-nonnegativity.} The operator products described above have a non-negative expectation value also if they appear \textit{sandwiched} between another product of operators and its conjugate.
Indeed, if $O_2$ is a product of operators that has non-negative expectation value with any state, then $\braket{\psi|O_1^\dagger O_2 O_1|\psi}=\braket{\psi'|O_2|\psi'}\geq 0$ for any $O_1$, $\ket{\psi}$.
An example of these products are those that correspond to subsequent measurements on a same state, whose expectation values represent the probabilities of sequences of outcomes.
This feature is also automatically imposed when generating an instance of \texttt{InflationSDP}.

\paragraph{Linearized polynomial constraints.} Both in the inflation methods for compatibility with quantum (based on semidefinite programming) and classical and generalized physical models (based on linear programming), it is possible to transform certain non-linear relations between the unknowns in the problems into linear ones when one assesses the compatibility of a given distribution with the scenario.
When a subgraph of the inflation contains more than one connected components, and (at least) one of these components can be associated a numerical value from the distribution under scrutiny (these are known as injectable components in the terminology of Ref.~\cite{wolfe2019}), the variables associated to the subgraph can be related to those corresponding to its non-injectable connected components via linear relations, reducing the feasible region.
Linearized polynomial constraints are used, for instance, in Ref.~\cite{pozas2022proofs} in the case of inflation for compatibility with classical models.
For compatibility with quantum models, \cite[App. D.2]{AlexThesis} contains a demonstration of its advantage in several scenarios.
It should be noted that when using linearized polynomial constraints, in case of infeasibility, the dual of the semidefinite (or linear) program can only serve as a certificate of infeasibility for the tested distribution, given that for other distributions the feasible region reduction under linearized polynomial constraints can be different \cite{pozas2022proofs}.
\ourpackagename{} allows to impose linearized polynomial constraints by setting the flag \texttt{use\_lpi\_constraints=True} in \texttt{set\_values()}. 

\section{Main functionality}
\label{sec:main}
We showcase here the user experience in using \ourpackagename{} to solve a series of problems that routinely appear in causal inference scenarios. All these examples can be downloaded as ready-to-run Jupyter notebooks from the repository of the library (see Sec.~\ref{sec:additional:documentation}).

\subsection{Feasibility problems and extraction of certificates}
\label{sec:main:feas}
The first example we will consider is that of demonstrating that a particular distribution can not be generated in a specific network arrangement when the sources (recall, the latent nodes in the scenario characterized by the DAG) distribute quantum systems, and the parties (the visible nodes) perform quantum measurements in the systems received.
We illustrate this type of problems by showing that the so-called W distribution, defined as 
\begin{equation}
    P_W(a,b,c)= \begin{cases}\frac{1}{3} & \text { if } a+b+c=1, \\ 0 & \text { otherwise }\end{cases},
    \label{eq:W}
\end{equation}
where $a,b,c\in\{0,1\}$, is incompatible with the quantum triangle causal scenario of Fig.~\ref{fig:triangle}.

The first step is to specify the original scenario and its desired inflation by creating an instance of \texttt{InflationProblem}.
This requires specifying (i) the DAG of the original scenario as a dictionary where the keys are parent nodes and the values are lists of the corresponding children, (ii) the number of possible outcomes and number of possible measurement settings of each of the visible nodes in the scenario, and (iii) the inflation levels, which represent the number of copies of each latent node that will be considered in the inflated scenario: 
\begin{lstlisting}[language=Python, captionpos=t]
    dag = {"rho_AB": ["A", "B"],
           "rho_BC": ["B", "C"],
           "rho_AC": ["A", "C"]}
    nr_outputs = [2, 2, 2]
    nr_inputs  = [1, 1, 1]
    nr_copies  = [2, 2, 2]
    scenario   = InflationProblem(dag, nr_outputs, nr_inputs, nr_copies,
                                  order=['A', 'B', 'C'])
\end{lstlisting}
The order of the parties is specified via the \texttt{order} optional argument, and the order of the sources is taken to be the same as insertion order in the \texttt{dag} dictionary.

The generated instance of \texttt{InflationProblem} is fed to an \texttt{InflationSDP} instance.
This object controls all the features related to the semidefinite relaxation of the problem considered.
For instance, one can easily specify the relaxation obtained when using as generating monomials all products of operators of length at most 2 (this is, what is known as the second level in the NPA hierarchy \cite{npa1,npa2}):
\begin{lstlisting}[language=Python, captionpos=t]
    sdp = InflationSDP(scenario)
    sdp.generate_relaxation("npa2")
\end{lstlisting}

The last step is setting the constraints corresponding to observing the target probability distribution in marginals of the inflation distribution that can be identified with marginals of the original scenario.
If one wants to do this for all marginals, this can be achieved with the function \texttt{set\_distribution()}, but if more granularity is needed one must use the function \texttt{set\_values()} instead (see Sec.~\ref{sec:other:partial} for an explicit example using this function).
The distribution must be input as a multidimensional NumPy array \texttt{P[out${}_1$,...,out${}_n$,in${}_1$,...,in${}_n$]}, where each cell contains the corresponding probability, and where the $i$-th output and input, \texttt{out${}_i$} and \texttt{in${}_i$}, correspond to the party in the $i$-th position in the \texttt{order} keyword argument.
\begin{lstlisting}[language=Python, captionpos=t]
    sdp.set_distribution(P_W)
\end{lstlisting}

Then, running \texttt{sdp.solve()} executes the semidefinite program and stores its status in \texttt{sdp.status}.
For the problem at hand, this is \texttt{infeasible}, meaning that the W distribution of Eq.~\eqref{eq:W} does not admit a second-order (because of our specification of \texttt{nr\_copies}) quantum inflation of the triangle scenario, and thus by the inflation arguments (see Sec.~\ref{sec:inflation} and Ref.~\cite{wolfe2021}) it cannot be generated in the triangle causal scenario when the latent nodes represent sources of bipartite quantum states.

Once the problem is determined to be infeasible, semidefinite programming provides a certificate of such infeasibility.
These certificates of infeasibility, which witness incompatibility with a given inflation, can be transformed into polynomial Bell-like inequalities that witness distributions incompatible with the original causal scenario (see, e.g., \cite[Sec. VII.B]{wolfe2021}).
\ourpackagename{} automatically computes these certificates and provides them in a variety of useful forms.
For instance, running
\begin{lstlisting}[language=Python, captionpos=t]
    sdp.certificate_as_probs(clean=True)
\end{lstlisting}
produces as output a SymPy object of the following form
\[ -0.476 p_A(0|0)^2 + 0.059 p_A(0|0)p_{AB}(00|00) + 0.476 p_A(0|0)p_{ABC}(000|000) + \dots + 0.563,
\]
which signals distributions that produce a negative value as incompatible with the triangle scenario.
This object can be further manipulated easily with SymPy's built-in functions.

\paragraph{Feasibility as optimization.} In terms of numerical stability, it is often advised to frame feasibility problems as optimization problems.
This is specially relevant when the distribution whose compatibility we are testing is close to the boundary of the feasible set, where analytically feasible problems can be reported as infeasible. 

\ourpackagename{} allows for this framing by optimizing the smallest eigenvalue of the problem's moment matrix\footnote{If the largest value that the smallest eigenvalue can achieve is negative, it means that the moment matrix associated to the problem cannot be made positive-semidefinite, and thus the corresponding feasibility problem is infeasible.}.
This is achieved by passing one argument at solving time:
\begin{lstlisting}[language=Python, captionpos=t]
    sdp.solve(feas_as_optim=True)
\end{lstlisting}

The optimal (largest) value of the smallest eigenvalue of the moment matrix is stored in \texttt{sdp.objective\_value} and by inspecting its value one can determine whether the original problem is feasible (if $\text{\texttt{sdp.objective\_value}}\geq0$) or not (if \mbox{$\text{\texttt{sdp.objective\_value}}<0$}).
Furthermore, the quantity in \texttt{sdp.objective\_value} can provide a rudimentary notion of distance to the feasible set, and help estimating how the size of the feasible set changes when adding extra constraints to the problem or when changing the generating set.
Importantly, the extraction of certificates is unaffected, and can still be performed even when treating feasibility problems as optimization problems.

\paragraph{Physical moments as generating set.} It is known that, when considering the distribution in Eq.~\eqref{eq:W} subject to white noise,
\begin{equation} \label{eq:white_noise}
    \nu P_W + (1- \nu) \frac{1}{8}, \quad \nu\in[0,1],
\end{equation}
quantum inflation of order 2 can certify its incompatibility with the triangle with quantum latent nodes at least down to $\nu=0.8038$.
This result was obtained in Ref.~\cite{wolfe2021} by using a moment matrix of size 1175$\times$1175.
By using as monomials indexing the rows and columns of the moment matrix only those with non-negative expectation values under any quantum state as mentioned in Sec. \ref{sec:library}, one can recover the same $\nu_{\text{crit}}$ but with a much smaller moment matrix, of size 287$\times$287.
Due to its notable gains in memory (and its consequent gains in speed), using the non-negative monomials in the generating set is made very easy in \ourpackagename{}.
In order to recover the result mentioned above with the much smaller moment matrix, one just needs to run:
\begin{lstlisting}[language=Python, captionpos=t]
    genset = sdp.build_columns("physical2", max_monomial_length=4)
    sdp.generate_relaxation(genset)
    v = 0.8039
    sdp.set_distribution(v*P_W + (1-v)/8)
    sdp.solve()
    # infeasible
\end{lstlisting}
This improvement is most notable when dealing with problems that involve distributions without settings.
In our experience, in the case where the parties in the problem have a choice of different measurements to perform on the states received by the sources, gains are more moderate.

\subsection{Optimization of Bell operators} \label{sec:main:mermin}
The second large class of problems that can be solved within \ourpackagename{} is obtaining bounds on expectation values of Bell operators.
For example, let us consider Mermin's operator \cite{Mermin1990}:
$$ \text{Mermin} = \langle A_1 B_0 C_0 \rangle +  \langle A_0 B_1 C_0 \rangle +  \langle A_0 B_0 C_1 \rangle -  \langle A_1 B_1 C_1 \rangle, $$
where $\left\langle A_{x} B_{y} C_{z}\right\rangle =\sum_{a, b, c \in \{0,1\}} (-1)^{a+b+c}\, p(a,b,c|x,y,z)$.
Ref.~\cite{wolfe2021} bounds this quantity in the quantum triangle scenario by using its second-order inflation and a generating set composed of the union of the second level in the associated NPA hierarchy and the first local level, obtaining that its maximum value cannot be larger than 3.085.

In order to reproduce these results in \ourpackagename{}, one needs to pass the corresponding generating set to \texttt{generate\_relaxation()} and set the objective function.
The former can be easily done since \texttt{generate\_relaxation()} also admits an explicit list of operators as argument:
\begin{lstlisting}[language=Python, captionpos=t]
    npa2   = sdp.build_columns("npa2",   symbolic=True)
    local1 = sdp.build_columns("local1", symbolic=True)

    npa2_union_local1 = set(npa2).union(set(local1))
         
    sdp.generate_relaxation(list(npa2_union_local1))
\end{lstlisting}

Setting the objective is done via the function \texttt{set\_objective()}, which admits as input a SymPy object that is a polynomial of the operators involved in the problem
\begin{lstlisting}[language=Python, captionpos=t]
    mmnts = sdp.measurements
    A0, B0, C0, A1, B1, C1 = (1 - 2*mmnts[party][0][x][0] 
                                    for x     in [0, 1] 
                                    for party in [0, 1, 2])
    Mermin = A1*B0*C0 + A0*B1*C0 + A0*B0*C1 - A1*B1*C1
    
    sdp.set_objective(objective=Mermin, direction="max")
    sdp.solve()
    print(sdp.objective_value)
    # 3.0851...
\end{lstlisting}

Note that, as mentioned in \cite[Sec. VII.C.2]{wolfe2021}, one can also optimize polynomial expressions of distributions compatible with the original scenario as long as they can be written as linear combinations of products of the operators in the inflation.
For example, one can optimize the mean squared distance to a target distribution in a second-order inflation by using operators corresponding to non-overlapping inflation copies.
In \ourpackagename{}, these operators are stored in the second dimension of \texttt{sdp.measurements}.

\subsection{Bounds on critical parameter values}
A third large class of problems of interest in quantum information theory that can be solved in \ourpackagename{} is the calculation of critical values of some parameter that characterize the family of distributions under scrutiny.
Examples of this are the estimation of the maximum amount of noise \cite{scalarextension}, the maximum angle between the measurements of a party \cite{fullnn}, or the maximum probability of detection failure \cite{abiuso2022} beyond which nonlocality cannot be certified.
This type of problems can be handled within \ourpackagename{} by using the function \texttt{max\_within\_feasible}.
This function takes as input an instance of an \texttt{InflationSDP} that characterizes the set of feasible distributions, and a mapping (in the form of a Python dictionary) from cells in the corresponding moment matrix to arbitrary symbolic expressions depending on the variable to be optimized.

Currently \ourpackagename{} features two ways for obtaining critical parameter values, which are specified by setting the corresponding flag in \texttt{max\_within\_feasible}.
The first one, \texttt{method="bisection"}, is a bisection algorithm, which takes increasingly small steps in the direction of the critical value of the parameter, taking $n=\lceil \log_2\Delta-\log_2\varepsilon\rceil$ iterations to reach a solution within accuracy $\varepsilon$ (where $\Delta$ is the width of the interval that the variable is constrained to lie in).
The second method, \texttt{method="dual"}, exploits the certificates of infeasibility in order to reduce the number of iterations required.
The certificates of infeasibility are surfaces that always leave the feasible region in the half-space that takes positive values.
This second method, instead of modifying the parameter by a fixed value, chooses as next candidate the value that lies in the boundary of the certificate, typically leading to fewer evaluations of semidefinite programs.
Furthermore, it is possible to use the functions \texttt{set\_values()} and \texttt{set\_distribution()} as shortcuts to obtain complete dictionaries of assignments that are stored in \texttt{InflationSDP.known\_moments}.

As an illustration of the use of \texttt{max\_within\_feasible}, the following example computes, after defining the relevant \texttt{InflationSDP}, the critical visibility for the W distribution of Eq.~\eqref{eq:W}:

\begin{lstlisting}[language=Python, captionpos=t]
    from inflation import max_within_feasible
    from sympy import Symbol
    v = Symbol("v")
    sdp.set_distribution(v*P_W + (1-v)/8)
    max_within_feasible(sdp, sdp.known_moments, "dual", bounds=[0, 1])
    # 0.8038...
\end{lstlisting}

\section{Further features}\label{sec:other}
In this section we collect additional functionality of \ourpackagename{} when restricting to specific kinds of problems and situations.
All the functionality described in this section can be combined seamlessly, both among them and with that described earlier.

\subsection{Characterization of classical causal models}
\label{sec:other:classical}
In quantum mechanics, operators corresponding to different parties commute, but this is not necessarily the case for operators corresponding to different measurements performed by a same party.
If different measurement operators for a same party commute, one can define a basis in which all operators are diagonal, hence obtaining results that can be recovered by measuring classical systems.
This means that, in analogy with works done in the multipartite Bell scenario~\cite{baccari2017}, imposing that all operators in a quantum inflation problem commute gives a relaxation of the set of correlations compatible with the given causal scenario where the latent nodes represent classical physical systems instead.
For a fixed inflation, the resulting NPA-like hierarchy converges to the linear program posed by the inflation technique for causal compatibility with classical latent nodes \cite{wolfe2019}, enabling the study of such problems with lower memory requirements. 

In order to restrict the characterizations generated to classical distributions in \ourpackagename{}, one just needs only to set the \texttt{commuting} flag to \texttt{True} when creating an instance of the \texttt{InflationSDP} object:
\begin{lstlisting}[language=Python, captionpos=t]
    sdp = InflationSDP(scenario, commuting=True)
\end{lstlisting}
The rest of the functions, such as for setting distributions and objective functions, or for extracting certificates, remains exactly the same. 

Given that, within quantum inflation, it is possible to consider compatibility with quantum and classical causal scenarios, a natural question is whether it is possible to assess compatibility with hybrid scenarios where some of the sources are classical and the remaining are quantum. It is indeed possible to consider such problems within the formalism of quantum inflation by enforcing commutations over certain subsets of operators. Analyzing hybrid scenarios is a feature not currently implemented in \ourpackagename{}, but that is expected to be supported in the future. We refer the reader to Sec.~\ref{sec:additional:future} for further details.

\subsection{Standard NPA hierarchy}
Since multipartite Bell scenarios (namely those where all parties receive states distributed by the same source) are particular instances of causal scenarios, \ourpackagename{} can also be used for analyzing standard multipartite quantum correlations using the NPA hierarchy \cite{npa1,npa2}.
In order to do so, one can call \texttt{InflationProblem} without specifying its \texttt{dag} argument.
For instance, the code to optimize the CHSH inequality in the bipartite Bell scenario is:
\begin{lstlisting}[language=Python, captionpos=t]
    scenario = InflationProblem(outcomes_per_party=[2, 2],
                                settings_per_party=[2, 2])
    # UserWarning: The DAG must be a non-empty dictionary with parent variables as keys and lists of children as values. Defaulting to one global source.
    sdp = InflationSDP(scenario)
    sdp.generate_relaxation("npa1")
    mmnts = sdp.measurements
    A0, B0, A1, B1 = (1 - 2*mmnts[party][0][x][0] for x     in [0, 1] 
                                                  for party in [0, 1])
    CHSH = A0*(B0+B1) + A1*(B0-B1)
    sdp.set_objective(CHSH)
    sdp.solve()
    print(sdp.objective_value)
    # 2.8284...
\end{lstlisting}

\subsection{Scenarios with partial information}
\label{sec:other:partial}
\ourpackagename{} can also handle scenarios where not all the information about a particular distribution in the original scenario is known.
An important example is the analysis of cryptographic scenarios, where the honest parties may know what is their joint distribution but they can not know what is the joint distribution with a potential adversary.
One simple such scenario is considered in Ref.~\cite[Sec. VIII]{wolfe2021}.
Specifying particular elements of a distribution in an \texttt{InflationSDP} object is achieved via the use of the function \texttt{set\_values()}, which admits as input a dictionary where the keys are the variables to be assigned numerical quantities, and the corresponding values are the quantities themselves.
In order to address the problem in Ref.~\cite[Sec. VIII]{wolfe2021} in \ourpackagename{}, one would write:

\begin{lstlisting}[language=Python, captionpos=t]
    p0 = sum(probability[0,b,0,0,0] for b in range(4))
    mmnts = sdp.measurements
    A = mmnts[0][0]
    B = mmnts[0][0][0]
    C = mmnts[2][0]
    E = mmnts[3][0][0]
    sdp.set_objective(A[0][0]*C[0][0]*E[0]/p0 - E[0])
    known_values = {}
    for a, b, c, x, z in numpy.ndindex(1, 3, 1, 2, 2):
        known_values[A[x][a]*B[b]*C[z][c]] = probability[a, b, c, x, z]
    # Proceed analogously for all other known quantities
    sdp.set_values(known_values)
\end{lstlisting}

\subsection{Scenarios beyond networks} \label{sec:other:instrumental}
\begin{figure}
    \centering
    \subfigure[\label{fig:instrumental:orig}]
    {\begin{minipage}[t]{0.4\textwidth}
          \centering\includegraphics[width=0.9\textwidth]{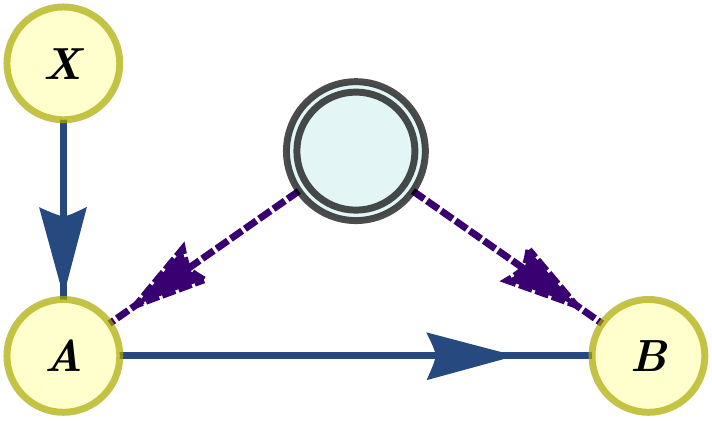}
      \end{minipage}
      }
      \hspace*{0.15cm}
    \subfigure[\label{fig:instrumental:network}]
    {\begin{minipage}[t]{0.4\textwidth}
          \centering\includegraphics[width=0.9\linewidth]{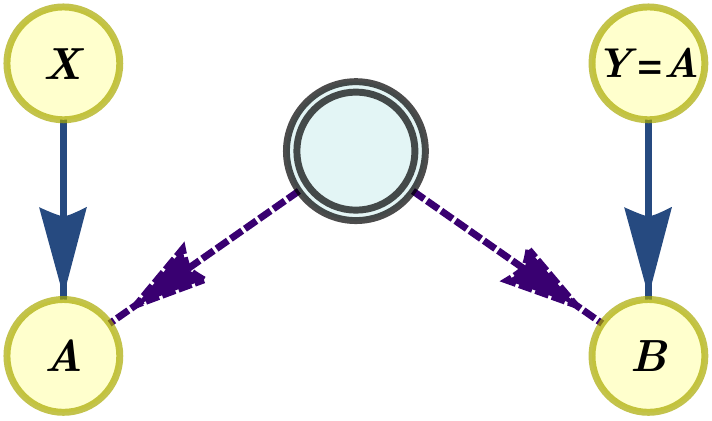}
      \end{minipage}
      }
    \caption{\protect{\subref{fig:instrumental:orig}} The instrumental scenario written as a DAG.
    Because of the arrow originating at $A$ and pointing to $B$, the scenario is not a network.
    \protect{\subref{fig:instrumental:network}} The interruption of \protect{\subref{fig:instrumental:orig}}. This is a network scenario, where the input to $B$ and the output of $A$ are restricted to coincide.
    The similarity of this scenario with the Bell scenario (recall Fig.~\ref{fig:bell}) has motivated its investigation within quantum information theory.}
    \label{fig:instrumental}
\end{figure}

So far, all the examples described have involved causal scenarios known as networks.
These are bipartite DAGs with a layer of visible variables denoting the parties outcomes, and a layer of both latent and visible variables that denote the sources of physical systems and the measurements performed by the parties on them, respectively.
Importantly, there are no connections between the nodes in each of the layers.
Not all DAGs fall in this category, and some non-network DAGs have received considerable attention recently in the literature.
The most important example is the so-called instrumental scenario~\cite{causalBook}, which has been extensively studied in the quantum information literature \cite{vanhimbeeck2019,gachechiladze2020,agresti2020,agresti2022} (see Fig.~\ref{fig:instrumental:orig}).
\ourpackagename{} is capable of handling with causal inference problems in these scenarios, by internally considering equivalent network-type scenarios such as that depicted in Fig.~\ref{fig:instrumental:network} (see \cite{navascues2020,wolfe2021} for the details on the equivalence).
The user experience in considering these problems is no different to that of considering causal inference over network-type DAGs.
As an illustration, the following snippet recovers the bounds of Bonet's inequality in \cite[Eq. (23)]{vanhimbeeck2019}.

\begin{lstlisting}[language=Python, captionpos=t]
    from sympy import Symbol as Sym
    inst = InflationProblem(dag={"rhoAB": ["A", "B"],
                                 "A": ["B"]},
                            outcomes_per_party=(2, 2),
                            settings_per_party=(3, 1))
    sdp = InflationSDP(inst)
    sdp.generate_relaxation("local1")
    objective = Sym("pAB(00|00)")
    objective += Sym("pA(1|0)")    - Sym("pAB(10|00)")  # pAB(11|0) 
    objective += Sym("pAB(00|10)") + Sym("pAB(10|10)")  # pB(0|1)
    objective += Sym("pA(0|2)")    - Sym("pAB(00|20)")  # pAB(01|2)
    sdp.set_objective(objective)
    sdp.solve()
    # 2.2071...
\end{lstlisting}

\subsection{Feasibility based on distribution supports}
The violation of Bell-type inequalities is the main method for the certification of non-classical behavior.
However, there exist even simpler certifications of non-classicality, that instead rely on possibilistic arguments: instead of setting bounds on combinations of the elements of compatible probability distributions, they only assume whether certain events are possible (positive probability) or impossible (zero probability) in order to reach a contradiction.
These certificates are known as Hardy-type paradoxes \cite{Mansfield_2012}.

\ourpackagename{} can handle proofs of non-classicality and non-quantumness in arbitrary DAGs based on possibilistic arguments.
It does so by assessing whether a quantum inflation (with commuting or non-commuting operators, respectively) exists where the probability elements inside the support are constrained to lie in the interval $[1,\infty)$ while those outside the support are given the value $0$.
This represents a re-scaled version of a standard quantum inflation moment matrix, $\Gamma^*=\Gamma/\epsilon$, that does not suffer of floating-point instabilities when determining if a probability element is outside the support or has assigned a very small value.

In order to deal with possibilistic feasibility problems in \ourpackagename{} one sets the argument \texttt{supports\_problem=True} when instantiating \texttt{InflationSDP}.
As an example, in order to recover the result that that the distribution from \cite[Eq.\,(19)]{vanhimbeeck2019} is incompatible with the scenario of Fig.~\ref{fig:instrumental:orig}, one would run the following code.
\begin{lstlisting}[language=Python, captionpos=t]
    sdp = InflationSDP(inst, supports_problem=True)
    sdp.generate_relaxation("local1")
    sdp.set_distribution(P_Eq19)
    sdp.solve()
    # "infeasible"
\end{lstlisting}
Note that, in contrast with other functionalities discussed in this section, optimization of objective functions is not possible when assessing the feasibility of distribution supports.

\section{Additional library information}
\label{sec:additional}
Here we provide further information concerning the development of the \ourpackagename{} library.

\subsection{Computational considerations}
\ourpackagename{} has been developed with speed and efficiency in mind.
It uses just-in-time compilation through Numba \cite{numba} in order to speed up core calculations, and dictionary caching to avoid needless function calls.
This results in all examples in the documentation being executable on a standard laptop with 8 GB of RAM.
Moment matrices of around $200$ columns and $1 500$ free scalar variables can be generated in 5 seconds and the SDP solved in 3 seconds; those of around $2 000$ columns and around $2 500$ variables can be generated in 8 minutes and the SDP solved in 10 minutes, and those of around $20 000$ columns can be generated in 17 hours.
In this last case, however, the SDP is too large to be solvable on a traditional desktop computer.\footnote{Tested on a PC with an Intel Core i9-10900X CPU @ 3.70GHz and 32 GB of RAM.} In order to solve these larger problems, a promising venue to pursue is using symmetries to block diagonalize the moment matrix. In the \href{https://ecboghiu.github.io/inflation/_build/html/advanced.html}{Documentation} we provide an example using the MATLAB software \texttt{RepLAB} \cite{Rosset21}.

\subsection{Future extensions} \label{sec:additional:future}
Inflation is a general technique that comprises a collection of routines specific to different types of physical systems.
Moreover, the fact that it is a young technique makes it not unreasonable to expect that refinements and alternative hierarchies will be developed in the future.
For these reasons, \ourpackagename{} is built having modularity in mind, so that new functionalities are easy to implement.

The main feature of the current implementation of \ourpackagename{} is the characterization of sets of quantum correlations in networks and certain non-network causal structures.
Moreover, it is also capable of handling the characterization of sets of classical correlations, and it contains the necessary equipment to handle simple non-network scenarios.
Subsequent releases will be focused on consolidating these capabilities, as well as adding functionalities to improve user experience and increase the range of problems and scenarios it can handle.
Planned additions to the library include:

\paragraph{Arbitrary causal scenarios.} One of the most exciting features of inflation methods is their ability to handle scenarios with latent-to-latent, visible-to-visible and visible-to-latent connections.
By the application of unpacking and exogenization algorithms (see Refs.~\cite{navascues2020,wolfe2021} for their descriptions), probability distributions compatible with any causal structure encoded in a DAG can be transformed into and analyzed as distributions compatible with an equivalent network-form DAG and satisfying additional equality constraints.
While the library already allows handling scenarios with visible-to-visible connections, in future versions we plan to add support for scenarios with visible-to-latent connections.
The automatic handling of this type of networks will mostly be integrated in the \texttt{InflationProblem} object, although it is expected that, due to the need of handling differently the cases of classical and quantum latent nodes (see \cite[Fig. 8]{wolfe2021}), processing is needed also further down in the pipeline.

\paragraph{Inflation based on linear programming.} As mentioned in Sec.~\ref{sec:other:classical}, the current implementation of \ourpackagename{} allows for the characterization of sets of classical probability distributions in network scenarios by means of semidefinite relaxations involving commuting operators, in the spirit of Ref.~\cite{baccari2017}.
A more direct way of performing such characterization is implementing the classical inflation hierarchy described in Refs.~\cite{wolfe2019,navascues2020}, based on linear programming.
Fulfilment of this task involves the creation of a new fundamental object in the library, \texttt{InflationLP}, that will moreover be used for the characterization of distributions generated by measurements on systems described by generalized probabilistic theories (note that these distributions are a superset of the quantum distributions, and cannot be analyzed using quantum inflation).

\paragraph{Hybrid scenarios.} Most research in the field investigating correlations in networks has thus far focused on networks in which all sources and operations are described by the same physical theory, be it classical mechanics, quantum mechanics or a generalized probabilistic theory.
Studying the correlations that can arise in networks where different sources (and corresponding measurements) are described by different physical models is therefore an interesting question, with consequences both at the theoretical~\cite{fullnn} and at the practical levels.
Subsequent versions of the library will admit the specification of whether each source in the scenario is classical or quantum, and implement these using the fact that measurements on classical systems only reveal pre-existing properties and the order in which these are revealed is unimportant.
The commutation relations between the operators in a scenario with classical and quantum sources are thus as follows: when a party receives systems only from classical sources, then all operators associated with that party commute with each other, including operators with different settings or acting on partially-overlapping sets of systems.
For a more generic party, connected to some number (including zero) of classical sources as well as some number of quantum sources, the commutation rules are the following: if two operators act on non-overlapping sets of states, they commute as usual, since these are operators acting on different Hilbert spaces; if they act on exactly the same set of states, then they commute only if they are identical or orthogonal (i.e., referring to the same setting); if they act on sets of states with partial overlap, then they commute only if all the states in the overlap come from classical sources, since this represents the situation where the party measures different quantum systems and the same classical systems; otherwise (i.e., if the operators act on sets of states with partial overlap, and any of the states in the overlap is quantum), the operators do not commute, even if they refer to same settings and/or outcomes.

\paragraph{Interfacing.} Future plans include adding support for other optimizers widely available such as SDPT3~\cite{sdpt3}, CVXPY~\cite{diamond2016cvxpy}, SCS~\cite{scs} and Gurobi~\cite{gurobi} (this last one is especially interesting since it is not restricted to linear and semidefinite programming problems), and translating the problems generated to forms compatible with other optimization libraries such as PICOS~\cite{picos} and CVXOPT~\cite{cvxopt}.
Interfacing with other tools, such as \texttt{SDPSymmetryReduction.jl}~\cite{brosch2020jordan} for reducing the memory and computational load of the problems via exploitation of symmetries, and scalar extension~\cite{scalarextension} for imposing the independence of variables in the inflation scenarios, will also prove useful.
Currently, it is possible to export the problem in a MATLAB-compatible form that can be directly read out by \texttt{RepLAB}~\cite{Rosset21} in order to block diagonalize it.

\subsection{Documentation for \ourpackagename{}}
\label{sec:additional:documentation}
The documentation of the library, which includes a user's guide and an API glossary, can be found online at \href{https://ecboghiu.github.io/inflation}{https://ecboghiu.github.io/inflation}.
The user’s guide contains more information on the installation and the topics covered in this manuscript, as well as subjects not covered here; for example, more applications, tips on improving performance, and in-depth tutorials.
The API glossary is automatically generated from the documentation comments written in the code and contains information about the public functions and classes defined.

\subsection{Contribution guidelines}
We welcome contributions to \ourpackagename{} from the larger community interested in causality, quantum nonlocality, and software for quantum information theory.
Contributions can come in the form of feedback about the library, feature requests, bug fixes, or code contributions (pull requests).
Feedback and feature requests can be done by opening an issue on the \href{https://www.github.com/ecboghiu/inflation}{\ourpackagename{} GitHub repository}.
Bug fixes and other pull requests can be done by forking the \ourpackagename{} source code, making changes, and then opening a pull request to the \ourpackagename{} GitHub repository.
Pull requests are peer-reviewed by \ourpackagename{}'s core developers to provide feedback and/or request changes.

Contributors are expected to adhere to \ourpackagename{} development practices including style guidelines and unit tests.
Tests are written with the UnitTest Python framework and are implemented outside the module.
To test installation or changes, one can download the source code from the repository, and use standard UnitTest functions.
For example, executing the following in a Unix terminal in the test folder runs all the tests: 
\begin{lstlisting}[language=Python, captionpos=t]
    python -m unittest -v
\end{lstlisting}

More details can be found in the \href{https://github.com/ecboghiu/inflation/blob/main/CONTRIBUTE.md}{Contribution guidelines documentation}.

\section{Concluding remarks}
\label{sec:conclusions}
We have presented the first open-source implementation of the inflation framework for causal compatibility.
Its focus is put in user experience and modularity, with the goal of being easy to use off the shelf while allowing for modifications needed by expert users.

While the current core implements the quantum inflation technique of Ref.~\cite{wolfe2021}, the library can be used off the shelf to characterize the sets of classical and quantum correlations in any network-type DAG, and non-network scenarios with visible-to-visible connections.
After briefly mentioning the principles behind inflation, we described the main components of the library and techniques to achieve tighter characterizations, and we illustrated its use in multiple problems of interest.
Finally, we outlined different ways in which the library can be extended to accommodate problems of interest for the broad community interested in quantum nonlocality and causality, and described additional software information including support and contribution guidelines.
We strongly encourage any willing user to contribute to the development of the library via \href{https://www.github.com/ecboghiu/inflation}{its repository}.

\section*{Acknowledgments}
We thank Sergio S\'anchez-Ram\'irez for discussions and comments.
This work is supported by the Spanish Ministry of Science and Innovation MCIN/AEI/10.13039/501100011033 (CEX2019-000904-S, CEX2019-000910-S, PRE2019-088482 and PID2020-113523GB-I00), the Spanish Ministry of Economic Affairs and Digital Transformation (project QUANTUM ENIA, as part of the Recovery, Transformation and Resilience Plan, funded by EU program NextGenerationEU), Comunidad de Madrid (QUITEMAD-CM P2018/TCS-4342), the CSIC Quantum Technologies Platform PTI-001, Fundació Cellex, Fundació Mir-Puig, and Generalitat de Catalunya through the CERCA program.
Research at Perimeter Institute is supported in part by the Government of Canada through the Department of Innovation, Science and Economic Development and by the Province of Ontario through the Ministry of Colleges and Universities.
\bibliographystyle{quantum}
\bibliography{biblio}

\begin{thebibliography}{10}

\bibitem{causalBook}
Judea Pearl.
\newblock ``{Causality: Models, Reasoning, and Inference}''.
\newblock \href{https://dx.doi.org/10.1017/CBO9780511803161}{Cambridge
  University Press}. ~(2009).

\bibitem{geiger1995}
Dan Geiger and Christopher Meek.
\newblock ``Quantifier elimination for statistical problems''.
\newblock In Proc. 15th Conf. Uncert. Artif. Intell. (AUAI, 1999).
\newblock \href{https://dx.doi.org/10.48550/arXiv.1301.6698}{Page 226–235}.
\newblock ~(1995).
\newblock  \href{http://arxiv.org/abs/1301.6698}{arXiv:1301.6698}.

\bibitem{tian2002}
Jin Tian and Judea Pearl.
\newblock ``On the testable implications of causal models with hidden
  variables''.
\newblock In Proc. 18th Conf. Uncert. Artif. Intell. (AUAI, 2002).
\newblock \href{https://dx.doi.org/10.48550/arXiv.1301.0608}{Page 519–527}.
\newblock ~(2002).
\newblock  \href{http://arxiv.org/abs/1301.0608}{arXiv:1301.0608}.

\bibitem{garcia2005}
Luis~David Garcia, Michael Stillman, and Bernd Sturmfels.
\newblock ``Algebraic geometry of {B}ayesian networks''.
\newblock \href{https://dx.doi.org/10.1016/j.jsc.2004.11.007}{J. Symb. Comput.
  {\bf 39}, 331--355}~(2005).
\newblock  \href{http://arxiv.org/abs/math/0301255}{arXiv:math/0301255}.

\bibitem{garcia2014}
Luis~David Garcia.
\newblock ``Algebraic statistics in model selection''.
\newblock In Proc. 20th Conf. Uncert. Artif. Intell. (AUAI, 2004).
\newblock \href{https://dx.doi.org/10.48550/arXiv.1207.4112}{Page 177–184}.
\newblock ~(2014).
\newblock  \href{http://arxiv.org/abs/1207.4112}{arXiv:1207.4112}.

\bibitem{lee2017}
Ciarán~M. Lee and Robert~W. Spekkens.
\newblock ``Causal inference via algebraic geometry: Feasibility tests for
  functional causal structures with two binary observed variables''.
\newblock \href{https://dx.doi.org/10.1515/jci-2016-0013}{J. Causal Inference
  {\bf 5}, 20160013}~(2017).
\newblock  \href{http://arxiv.org/abs/1506.03880}{arXiv:1506.03880}.

\bibitem{BellReview}
Nicolas Brunner, Daniel Cavalcanti, Stefano Pironio, Valerio Scarani, and
  Stephanie Wehner.
\newblock ``Bell nonlocality''.
\newblock \href{https://dx.doi.org/10.1103/RevModPhys.86.419}{Rev. Mod. Phys.
  {\bf 86}, 419--478}~(2014).
\newblock  \href{http://arxiv.org/abs/1303.2849}{arXiv:1303.2849}.

\bibitem{BellTheorem}
John~S. Bell.
\newblock ``On the {Einstein-Podolsky-Rosen} paradox''.
\newblock \href{https://dx.doi.org/10.1103/PhysicsPhysiqueFizika.1.195}{Physics
  Physique Fizika {\bf 1}, 195--200}~(1964).

\bibitem{wood2015}
Christopher~J. Wood and Robert~W. Spekkens.
\newblock ``The lesson of causal discovery algorithms for quantum correlations:
  causal explanations of {B}ell-inequality violations require fine-tuning''.
\newblock \href{https://dx.doi.org/10.1088/1367-2630/17/3/033002}{New J. Phys.
  {\bf 17}, 033002}~(2015).
\newblock  \href{http://arxiv.org/abs/1208.4119}{arXiv:1208.4119}.

\bibitem{chaves2015}
Rafael Chaves, Richard Kueng, Jonatan~B. Brask, and David Gross.
\newblock ``Unifying framework for relaxations of the causal assumptions in
  {B}ell's theorem''.
\newblock \href{https://dx.doi.org/10.1103/PhysRevLett.114.140403}{Phys. Rev.
  Lett. {\bf 114}, 140403}~(2015).
\newblock  \href{http://arxiv.org/abs/1411.4648}{arXiv:1411.4648}.

\bibitem{bilocal_branciard_prl}
Cyril Branciard, Nicolas Gisin, and Stefano Pironio.
\newblock ``Characterizing the nonlocal correlations created via entanglement
  swapping''.
\newblock \href{https://dx.doi.org/10.1103/PhysRevLett.104.170401}{Phys. Rev.
  Lett. {\bf 104}, 170401}~(2010).
\newblock  \href{http://arxiv.org/abs/0911.1314}{arXiv:0911.1314}.

\bibitem{bilocal_branciard_pra}
Cyril Branciard, Denis Rosset, Nicolas Gisin, and Stefano Pironio.
\newblock ``Bilocal versus nonbilocal correlations in entanglement-swapping
  experiments''.
\newblock \href{https://dx.doi.org/10.1103/PhysRevA.85.032119}{Phys. Rev. A
  {\bf 85}, 032119}~(2012).
\newblock  \href{http://arxiv.org/abs/1112.4502}{arXiv:1112.4502}.

\bibitem{Fritz_2012}
Tobias Fritz.
\newblock ``Beyond {B}ell{\textquotesingle}s theorem: correlation scenarios''.
\newblock \href{https://dx.doi.org/10.1088/1367-2630/14/10/103001}{New J. Phys.
  {\bf 14}, 103001}~(2012).
\newblock  \href{http://arxiv.org/abs/1206.5115}{arXiv:1206.5115}.

\bibitem{fraser2018}
Thomas~C. Fraser and Elie Wolfe.
\newblock ``Causal compatibility inequalities admitting quantum violations in
  the triangle structure''.
\newblock \href{https://dx.doi.org/10.1103/PhysRevA.98.022113}{Phys. Rev. A
  {\bf 98}, 022113}~(2018).
\newblock  \href{http://arxiv.org/abs/1709.06242}{arXiv:1709.06242}.

\bibitem{vanhimbeeck2019}
Thomas van Himbeeck, Jonatan Bohr~Brask, Stefano Pironio, Ravishankar
  Ramanathan, Ana~Bel{\'{e}}n Sainz, and Elie Wolfe.
\newblock ``Quantum violations in the {I}nstrumental scenario and their
  relations to the {B}ell scenario''.
\newblock \href{https://dx.doi.org/10.22331/q-2019-09-16-186}{{Quantum} {\bf
  3}, 186}~(2019).
\newblock  \href{http://arxiv.org/abs/1804.04119}{arXiv:1804.04119}.

\bibitem{networkReview}
Armin Tavakoli, Alejandro Pozas-Kerstjens, Ming-Xing Luo, and Marc-Olivier
  Renou.
\newblock ``Bell nonlocality in networks''.
\newblock \href{https://dx.doi.org/10.1088/1361-6633/ac41bb}{Rep. Prog. Phys.
  {\bf 85}, 056001}~(2022).
\newblock  \href{http://arxiv.org/abs/2104.10700}{arXiv:2104.10700}.

\bibitem{scalarextension}
Alejandro Pozas-Kerstjens, Rafael Rabelo, {\L}ukasz Rudnicki, Rafael Chaves,
  Daniel Cavalcanti, Miguel Navascu\'es, and Antonio Ac\'{\i}n.
\newblock ``Bounding the sets of classical and quantum correlations in
  networks''.
\newblock \href{https://dx.doi.org/10.1103/PhysRevLett.123.140503}{Phys. Rev.
  Lett. {\bf 123}, 140503}~(2019).
\newblock  \href{http://arxiv.org/abs/1904.08943}{arXiv:1904.08943}.

\bibitem{covariancematrices1}
Aditya Kela, Kai {Von Prillwitz}, Johan {\AA{}berg}, Rafael {Chaves}, and David
  {Gross}.
\newblock ``Semidefinite tests for latent causal structures''.
\newblock \href{https://dx.doi.org/10.1109/TIT.2019.2935755}{IEEE Trans. Inf.
  Theory {\bf 66}, 339--349}~(2020).
\newblock  \href{http://arxiv.org/abs/1701.00652}{arXiv:1701.00652}.

\bibitem{covariancematrices2}
Johan \AA{}berg, Ranieri Nery, Cristhiano Duarte, and Rafael Chaves.
\newblock ``Semidefinite tests for quantum network topologies''.
\newblock \href{https://dx.doi.org/10.1103/PhysRevLett.125.110505}{Phys. Rev.
  Lett. {\bf 125}, 110505}~(2020).
\newblock  \href{http://arxiv.org/abs/2002.05801}{arXiv:2002.05801}.

\bibitem{luo2018}
Ming-Xing Luo.
\newblock ``Computationally efficient nonlinear {Bell} inequalities for quantum
  networks''.
\newblock \href{https://dx.doi.org/10.1103/PhysRevLett.120.140402}{Phys. Rev.
  Lett. {\bf 120}, 140402}~(2018).
\newblock  \href{http://arxiv.org/abs/1707.09517}{arXiv:1707.09517}.

\bibitem{RenouFinner2019}
Marc-Olivier Renou, Yuyi Wang, Sadra Boreiri, Salman Beigi, Nicolas Gisin, and
  Nicolas Brunner.
\newblock ``Limits on correlations in networks for quantum and no-signaling
  resources''.
\newblock \href{https://dx.doi.org/10.1103/PhysRevLett.123.070403}{Phys. Rev.
  Lett. {\bf 123}, 070403}~(2019).
\newblock  \href{http://arxiv.org/abs/1901.08287}{arXiv:1901.08287}.

\bibitem{wolfe2019}
Elie Wolfe, Robert~W. Spekkens, and Tobias Fritz.
\newblock ``The inflation technique for causal inference with latent
  variables''.
\newblock \href{https://dx.doi.org/10.1515/jci-2017-0020}{J. Causal Inference
  {\bf 7}, 20170020}~(2019).
\newblock  \href{http://arxiv.org/abs/1609.00672}{arXiv:1609.00672}.

\bibitem{wolfe2021}
Elie Wolfe, Alejandro Pozas-Kerstjens, Matan Grinberg, Denis Rosset, Antonio
  Ac\'{\i}n, and Miguel Navascu\'es.
\newblock ``Quantum inflation: A general approach to quantum causal
  compatibility''.
\newblock \href{https://dx.doi.org/10.1103/PhysRevX.11.021043}{Phys. Rev. X
  {\bf 11}, 021043}~(2021).
\newblock  \href{http://arxiv.org/abs/1909.10519}{arXiv:1909.10519}.

\bibitem{gisin2020}
Nicolas Gisin, Jean-Daniel Bancal, Yu~Cai, Patrick Remy, Armin Tavakoli,
  Emmanuel Zambrini~Cruzeiro, Sandu Popescu, and Nicolas Brunner.
\newblock ``Constraints on nonlocality in networks from no-signaling and
  independence''.
\newblock \href{https://dx.doi.org/10.1038/s41467-020-16137-4}{Nat. Commun.
  {\bf 11}, 2378}~(2020).
\newblock  \href{http://arxiv.org/abs/1906.06495}{arXiv:1906.06495}.

\bibitem{fullnn}
Alejandro Pozas-Kerstjens, Nicolas Gisin, and Armin Tavakoli.
\newblock ``Full network nonlocality''.
\newblock \href{https://dx.doi.org/10.1103/PhysRevLett.128.010403}{Phys. Rev.
  Lett. {\bf 128}, 010403}~(2022).
\newblock  \href{http://arxiv.org/abs/2105.09325}{arXiv:2105.09325}.

\bibitem{pozas2022proofs}
Alejandro Pozas-Kerstjens, Nicolas Gisin, and Marc-Olivier Renou.
\newblock ``Proofs of network quantum nonlocality in continuous families of
  distributions''.
\newblock \href{https://dx.doi.org/10.1103/PhysRevLett.130.090201}{Phys. Rev.
  Lett. {\bf 130}, 090201}~(2023).
\newblock  \href{http://arxiv.org/abs/2203.16543}{arXiv:2203.16543}.

\bibitem{code}
Emanuel-Cristian Boghiu, Elie Wolfe, and Alejandro Pozas-Kerstjens.
\newblock ``Source code for \texttt{inflation}''.
  \href{https://doi.org/10.5281/zenodo.7305544}{Zenodo \textbf{7305544}}
  (2022).

\bibitem{baccari2017}
Flavio Baccari, Daniel Cavalcanti, Peter Wittek, and Antonio Ac\'{\i}n.
\newblock ``Efficient device-independent entanglement detection for
  multipartite systems''.
\newblock \href{https://dx.doi.org/10.1103/PhysRevX.7.021042}{Phys. Rev. X {\bf
  7}, 021042}~(2017).
\newblock  \href{http://arxiv.org/abs/1612.08551}{arXiv:1612.08551}.

\bibitem{Steeg2011}
Greg ver Steeg and Aram Galstyan.
\newblock ``A sequence of relaxations constraining hidden variable models''.
\newblock In Proceedings of the Twenty-Seventh Conference on Uncertainty in
  Artificial Intelligence.
\newblock \href{https://dx.doi.org/10.48550/arXiv.1106.1636}{Page 717–726}.
\newblock UAI'11Arlington, Virginia, USA~(2011). AUAI Press.
\newblock  \href{http://arxiv.org/abs/1106.1636}{arXiv:1106.1636}.

\bibitem{navascues2020}
Miguel Navascu\'{e}s and Elie Wolfe.
\newblock ``The inflation technique completely solves the causal compatibility
  problem''.
\newblock \href{https://dx.doi.org/10.1515/jci-2018-0008}{J. Causal Inference
  {\bf 8}, 70 -- 91}~(2020).
\newblock  \href{http://arxiv.org/abs/1707.06476}{arXiv:1707.06476}.

\bibitem{ligthart2022}
Laurens~T. Ligthart and David Gross.
\newblock ``The inflation hierarchy and the polarization hierarchy are complete
  for the quantum bilocal scenario''~(2022).
\newblock  \href{http://arxiv.org/abs/2212.11299}{arXiv:2212.11299}.

\bibitem{ligthart2021}
Laurens~T. Ligthart, Mariami Gachechiladze, and David Gross.
\newblock ``A convergent inflation hierarchy for quantum causal
  structures''~(2021).
\newblock  \href{http://arxiv.org/abs/2110.14659}{arXiv:2110.14659}.

\bibitem{numpy}
Charles~R. Harris, K.~Jarrod Millman, St{\'{e}}fan~J. van~der Walt, et~al.
\newblock ``Array programming with {NumPy}''.
\newblock \href{https://dx.doi.org/10.1038/s41586-020-2649-2}{Nature {\bf 585},
  357--362}~(2020).

\bibitem{sympy}
Aaron Meurer, Christopher~P. Smith, Mateusz Paprocki, et~al.
\newblock ``{SymPy: symbolic computing in Python}''.
\newblock \href{https://dx.doi.org/10.7717/peerj-cs.103}{PeerJ Comput. Sci.
  {\bf 3}, e103}~(2017).

\bibitem{scipy}
Pauli Virtanen, Ralf Gommers, Travis~E. Oliphant, et~al.
\newblock ``{{SciPy} 1.0: Fundamental Algorithms for Scientific Computing in
  Python}''.
\newblock \href{https://dx.doi.org/10.1038/s41592-019-0686-2}{Nat. Methods {\bf
  17}, 261--272}~(2020).

\bibitem{numba}
Siu~Kwan Lam, Antoine Pitrou, and Stanley Seibert.
\newblock ``Numba: A {LLVM}-based {P}ython {JIT} compiler''.
\newblock In Proceedings of the Second Workshop on the LLVM Compiler
  Infrastructure in HPC.
\newblock \href{https://dx.doi.org/10.1145/2833157.2833162}{LLVM '15~}New York,
  NY, USA~(2015). Association for Computing Machinery.

\bibitem{mosek}
MOSEK ApS.
\newblock ``{MOSEK Fusion API for Python}''.
\newblock \url{https://docs.mosek.com/latest/pythonfusion/index.html}~(2019).

\bibitem{yalmip}
Johann L{\"{o}}fberg.
\newblock ``{YALMIP}: A toolbox for modeling and optimization in {MATLAB}''.
\newblock In Proceedings of the CACSD Conference.
\newblock Taipei, Taiwan~(2004).
\newblock  url:~\href{https://yalmip.github.io/}{yalmip.github.io/}.

\bibitem{npa1}
Miguel Navascu\'es, Stefano Pironio, and Antonio Ac\'{i}n.
\newblock ``Bounding the set of quantum correlations''.
\newblock \href{https://dx.doi.org/10.1103/PhysRevLett.98.010401}{Phys. Rev.
  Lett. {\bf 98}, 010401}~(2007).
\newblock
  \href{http://arxiv.org/abs/quant-ph/0607119}{arXiv:quant-ph/0607119}.

\bibitem{npa2}
Miguel Navascu\'es, Stefano Pironio, and Antonio Ac{\'i}n.
\newblock ``A convergent hierarchy of semidefinite programs characterizing the
  set of quantum correlations''.
\newblock \href{https://dx.doi.org/10.1088/1367-2630/10/7/073013}{New J. Phys.
  {\bf 10}, 073013}~(2008).
\newblock  \href{http://arxiv.org/abs/0803.4290}{arXiv:0803.4290}.

\bibitem{pna}
Stefano Pironio, Miguel Navascu\'es, and Antonio Ac{\'i}n.
\newblock ``Convergent relaxations of polynomial optimization problems with
  non-commuting variables''.
\newblock \href{https://dx.doi.org/10.1137/090760155}{SIAM J. Optim. {\bf 20},
  2157--2180}~(2010).
\newblock  \href{http://arxiv.org/abs/0903.4368}{arXiv:0903.4368}.

\bibitem{moroder2013}
Tobias Moroder, Jean-Daniel Bancal, Yeong-Cherng Liang, Martin Hofmann, and
  Otfried G\"uhne.
\newblock ``Device-independent entanglement quantification and related
  applications''.
\newblock \href{https://dx.doi.org/10.1103/PhysRevLett.111.030501}{Phys. Rev.
  Lett. {\bf 111}, 030501}~(2013).
\newblock  \href{http://arxiv.org/abs/1302.1336}{arXiv:1302.1336}.

\bibitem{AlexThesis}
Alejandro Pozas-Kerstjens.
\newblock ``Quantum information outside quantum information''.
\newblock PhD thesis.
\newblock Universitat Polit\'ecnica de Catalunya.
\newblock ~(2019).
\newblock
  url:~\href{http://hdl.handle.net/10803/667696}{http://hdl.handle.net/10803/667696}.

\bibitem{Mermin1990}
N.~David Mermin.
\newblock ``Quantum mysteries revisited''.
\newblock \href{https://dx.doi.org/10.1119/1.16503}{Amer. J. Phys. {\bf 58},
  731--734}~(1990).

\bibitem{abiuso2022}
Paolo Abiuso, Tam\'as Kriv\'achy, Emanuel-Cristian Boghiu, Marc-Olivier Renou,
  Alejandro Pozas-Kerstjens, and Antonio Ac\'{\i}n.
\newblock ``Single-photon nonlocality in quantum networks''.
\newblock \href{https://dx.doi.org/10.1103/PhysRevResearch.4.L012041}{Phys.
  Rev. Research {\bf 4}, L012041}~(2022).
\newblock  \href{http://arxiv.org/abs/2108.01726}{arXiv:2108.01726}.

\bibitem{gachechiladze2020}
Mariami Gachechiladze, Nikolai Miklin, and Rafael Chaves.
\newblock ``Quantifying causal influences in the presence of a quantum common
  cause''.
\newblock \href{https://dx.doi.org/10.1103/PhysRevLett.125.230401}{Phys. Rev.
  Lett. {\bf 125}, 230401}~(2020).
\newblock  \href{http://arxiv.org/abs/2007.01221}{arXiv:2007.01221}.

\bibitem{agresti2020}
Iris Agresti, Davide Poderini, Leonardo Guerini, Michele Mancusi, Gonzalo
  Carvacho, Leandro Aolita, Daniel Cavalcanti, Rafael Chaves, and Fabio
  Sciarrino.
\newblock ``Experimental device-independent certified randomness generation
  with an instrumental causal structure''.
\newblock \href{https://dx.doi.org/10.1038/s42005-020-0375-6}{Commun. Phys.
  {\bf 3}, 110}~(2020).
\newblock  \href{http://arxiv.org/abs/1905.02027}{arXiv:1905.02027}.

\bibitem{agresti2022}
Iris Agresti, Davide Poderini, Beatrice Polacchi, Nikolai Miklin, Mariami
  Gachechiladze, Alessia Suprano, Emanuele Polino, Giorgio Milani, Gonzalo
  Carvacho, Rafael Chaves, and Fabio Sciarrino.
\newblock ``Experimental test of quantum causal influences''.
\newblock \href{https://dx.doi.org/10.1126/sciadv.abm1515}{Sci. Adv. {\bf 8},
  eabm1515}~(2022).
\newblock  \href{http://arxiv.org/abs/2108.08926}{arXiv:2108.08926}.

\bibitem{Mansfield_2012}
Shane Mansfield and Tobias Fritz.
\newblock ``Hardy's non-locality paradox and possibilistic conditions for
  non-locality''.
\newblock \href{https://dx.doi.org/10.1007/s10701-012-9640-1}{Found. Phys. {\bf
  42}, 709--719}~(2012).
\newblock  \href{http://arxiv.org/abs/1105.1819}{arXiv:1105.1819}.

\bibitem{Rosset21}
Denis Rosset, Felipe Montealegre-Mora, and Jean-Daniel Bancal.
\newblock ``{RepLAB}: A computational/numerical approach to representation
  theory''.
\newblock In Quantum Theory and Symmetries.
\newblock \href{https://dx.doi.org/10.1007/978-3-030-55777-5_60}{Pages
  643--653}.
\newblock CRM Series in Mathematical Physics. Proceedings of the 11th
  International Symposium, Montreal,\,\,Springer~(2021).
\newblock  \href{http://arxiv.org/abs/1911.09154}{arXiv:1911.09154}.

\bibitem{sdpt3}
Kim-Chuan Toh, Michael~J. Todd, and Reha~H. Tütüncü.
\newblock ``{SDPT3} — a {MATLAB} software package for semidefinite
  programming''.
\newblock \href{https://dx.doi.org/10.1080/10556789908805762}{Optim. Metods
  Softw. {\bf 11}, 545--581}~(1999).

\bibitem{diamond2016cvxpy}
Steven Diamond and Stephen Boyd.
\newblock ``{CVXPY}: {A} {P}ython-embedded modeling language for convex
  optimization''.
\newblock \href{https://dx.doi.org/10.48550/arXiv.1603.00943}{J. Mach. Learn.
  Res. {\bf 17}, 1--5}~(2016).
\newblock  \href{http://arxiv.org/abs/1603.00943}{arXiv:1603.00943}.

\bibitem{scs}
Brendan O'Donoghue, Eric Chu, Neal Parikh, and Stephen Boyd.
\newblock ``{SCS: Splitting Conic Solver}''.
\newblock \url{https://github.com/cvxgrp/scs}~(2021).

\bibitem{gurobi}
{Gurobi Optimization, LLC}.
\newblock ``{Gurobi Optimizer Reference Manual}''.
\newblock \url{https://www.gurobi.com}~(2022).

\bibitem{picos}
Guillaume Sagnol and Maximilian Stahlberg.
\newblock ``{PICOS}: A {Python} interface to conic optimization solvers''.
\newblock \href{https://dx.doi.org/10.21105/joss.03915}{J. Open Source Softw.
  {\bf 7}, 3915}~(2022).

\bibitem{cvxopt}
Martin~S. Andersen, Joachim Dahl, and Lieven Vandenberghe.
\newblock ``{CVXOPT: Python software for convex optimization}''.
\newblock \url{http://cvxopt.org/}~(2015).

\bibitem{brosch2020jordan}
Daniel Brosch and Etienne de~Klerk.
\newblock ``Jordan symmetry reduction for conic optimization over the doubly
  nonnegative cone: theory and software''.
\newblock \href{https://dx.doi.org/10.1080/10556788.2021.2022146}{Optim.
  Methods Softw. {\bf 37}, 2001--2020}~(2022).
\newblock  \href{http://arxiv.org/abs/2001.11348}{arXiv:2001.11348}.

\end{thebibliography}

\end{document}